\documentclass[usletter, 10pt, conference]{cssconf}

\IEEEoverridecommandlockouts 

\usepackage{graphics}
\usepackage{amsmath} 
\usepackage{amssymb}  
\usepackage{mathptmx}

\usepackage{graphicx}
\usepackage{psfrag}

\newtheorem{thm}{Theorem}%
\newtheorem{lem}{Lemma}

\newtheorem{cor}{Corollary}

\newcommand{\RR}{{\mathbb R}}

\newcommand{\LLL}{{\mathcal  L}}

\title{On the inverse scattering of star-shape LC-networks
    \thanks{This work is partly supported by a PREDIT contract called SEEDS (Smart Embedded Electronic Diagnosis Systems),
    in cooperation between DELPHI, SERMA INGENIEURIE, Monditech, LGEP, INRIA and CEA LIST.}}

\author{  Filippo Visco Comandini, Mazyar Mirrahimi and Michel
Sorine
\thanks{The authors are with INRIA Rocquencourt,
        Domaine de Voluceau, B.P. 105, 78153 Le Chesnay cedex, France,
        {\tt\small filippo.comandini@inria.fr} and {\tt\small
        mazyar.mirrahimi@inria.fr} and {\tt\small
        michel.sorine@inria.fr}}
}

\begin{document}

\maketitle \thispagestyle{empty} \pagestyle{empty}

\begin{abstract}
The study of the scattering data for a star-shape network of
LC-transmission lines is transformed into the  scattering analysis
of a Schr\"odinger operator on the same graph. The boundary
conditions coming from the Kirchhoff rules ensure the existence of a
unique self-adjoint extension of the mentioned Schr\"odinger
operator. While the graph consists of a number of infinite branches
and a number finite ones, all joining at a central node, we provide
a construction of the scattering solutions. Under non-degenerate
circumstances (different wave travelling times for finite branches),
we show that the study of the reflection coefficient in the
high-frequency regime must provide us with the number of the
infinite branches as well as the the wave travelling times for
finite ones.
\end{abstract}

\section{Introduction}\label{sec:intro}

The number of electronic equipments is increasing rapidly in
automotive vehicles, aircrafts, and many other safety critical
systems. Consequently, the reliability of wired networks and
electric connections is becoming more and more important. For
example, in automotive industry, a goal is to develop compact and
easy to use devices for the diagnosis of electric connection
failures in garage or at the end of the production chain. These
devices should be capable of detecting and locating failures in
cables and in connectors. Another goal is on-board diagnosis: the
diagnosis device will be integrated to the vehicle in order to
detect failures under normal working conditions of the vehicle. To
find faulty wiring in such networks, it is not always possible to
measure end-to-end cable impedances, because the number of available
diagnostic port plugs is limited, and furthermore, for diagnosis
purpose, it is not sufficient to detect a high end-to-end impedance
as it is also necessary to locate the fault within the cable. In
such situations Time or Frequency Domain Reflectometry (TDR, FDR)
are the most commonly used methods: a high frequency signal (a short
travelling pulse for TDR, a standing wave for FDR) is sent down a
wire at some point and the signal reflected by the network is
measured at the same point and analyzed for fault detection and
location~\cite{Pan-et-al-02,vanhamme-90,furse-et-al-03,Auzanneau-et-al-07}.
Automatic fault detection and diagnosis using reflectometry methods
is the subject of intense research, both on the technologies of
"smart wiring systems" and reflectometers~\cite{sharma-et-al-07} and
on the foundations of the TDR/FDR methods. Most studies are
concerned with various aspects of mathematical modeling and
simulation of microwave propagation in networks, from network
complexity~\cite{parmantier-04} to reduced order modeling of high
frequency phenomena like skin effect~\cite{admane-et-al-07}. The
inverse problem of fault detection and localization for a network is
in general studied in simulation or experimentally. For this
problem, the available theory is developed mainly in the framework
of Inverse Scattering Theory limited to very simple networks
(segment, half-line, line) (see
e.g.~\cite{bruckstein-kailath-87,scattering-02}).

In this paper, we consider the inverse scattering problem (ISP) for
a star-shape network $\Gamma$ of transmission lines. Some of the
branches of the network are extended in a direction $z$ to $+\infty$
and the others admit a finite length. The network is assumed to be
non-dissipative; i.e. the series resistance $R(z)$ and the shunt
conductance $G(z)$ per unit length are assumed to vanish and
therefore, we are dealing with $LC$-wires.

\begin{figure}[h]
 \psfrag{r}{\tiny{$e_1$}}
 \psfrag{s}{\tiny{$e_2$}}
 \psfrag{p}{\tiny{$e_{m-1}$}}
 \psfrag{q}{\tiny{$e_{m}$}}
 \psfrag{a}{\tiny{$e_{m+n}$}}
 \psfrag{b}{\tiny{$e_{m+n-1}$}}
 \psfrag{c}{\tiny{$e_{m+2}$}}
 \psfrag{d}{\tiny{$e_{m+1}$}}
\begin{center}
\includegraphics[width=6 cm]{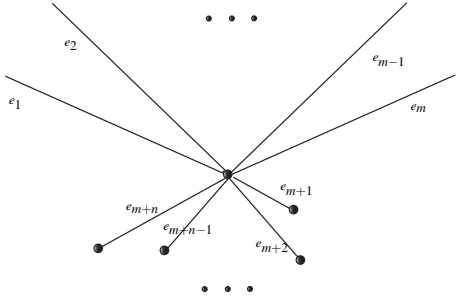}
\caption{The $LC$-network $\Gamma$ consisting of $m$ infinite
branches extended to $z=\infty$ and $n$ finite branches of various
lengthes.}\label{fig:graph}
\end{center}
\end{figure}

The main concern here is to derive some information concerning the
topology and the geometry of the network, applying a very few
scattering information. Indeed, fixing an infinite branch $e_1$ (see
Figure~\ref{fig:graph}), we generate a high-frequency time varying
voltage at its infinite end and measuring the intensity at the same
end, we would like to derive information such as the number of
infinite branches $m$ and the wave travelling time $\tau_{m+j}$ of
the finite segments $e_{m+j}$.

Here, in order to model our network, each branch is parameterized
through its own length. The central node is assumed to be the zero
(origin) of all the branches. While the position coordinate $z_j$
takes values inside the interval $[0,l_j]$, $l_j$ being the length
of the branch $e_j$ and therefore\small
$$
l_j=\infty\quad \forall j\in\{1,\cdots, m\}\quad \text{and}\quad
l_j<\infty\quad \forall j\in\{m+1,\cdots, m+n\}.
$$\normalsize

Moreover, on each branch, of finite or infinite length, we suppose
that
\begin{description}
\item[A1] $L(z)$ the inductance and $C(z)$ the capacitance are sufficiently regular (twice differentiable);
\item[A2] $L(z)>0$ and  $C(z)>0$.
\item[A3] On infinite branches, $L(z)$ and $C(z)$ have strictly positive
finite limits $L(\infty)$ and $C(\infty)$ as $z\rightarrow\infty$.
\end{description}

In the next section, we provide the mathematical model consisting of
the transmission line equations on each one of the network branches
and the boundary conditions (Kirchhoff rules) coupling these
equations together. Next, for the case of a simple transmission
line,  we provide a brief review of an old result by I.
Kay~\cite{kay-72}, transforming the inverse scattering problem for
one-dimensional nonuniform  non dissipative transmission lines to
the inverse scattering problem for a Schr\"odinger operator. We
extend this result to the case of our network by defining relevant
boundary conditions. Finally, we provide a result ensuring the
essential self-adjointness of the Schr\"odinger operator on the
graph with the aforementioned boundary conditions. The existence of
a unique self-adjoint extension allows us to start the study of the
scattering theory and the scattering solutions.

Next, in Section~\ref{sec:scat}, we solve the scattering problem by
providing the scattering solutions. In this aim, we apply the Jost
solutions on semi-infinite
branches~\cite{Gerasimenko-87,Gerasimenko-88} and the fundamental
solutions on the finite segments~\cite{marchenko-book}.

Finally, in section~\ref{sec:asy}, we study the asymptotic behavior
of the reflection coefficient in the limit of high-frequency waves
and we prove how this limit provides us the needed information on
the topology and the geometry of the graph.

\section{Mathematical model}\label{sec:math}
\subsection{Transmission line equations}

In the absence of dissipation, the transmission line equations
writes
\begin{equation}\label{Jaulent1}
\left\{\begin{matrix} \frac{\partial I}{\partial z}+ C(z) \frac{\partial U}{\partial t}=0\\
 \frac{\partial U}{\partial z}+ L(z) \frac{\partial I}{\partial t}=0\\
 \end{matrix}\right.
\end{equation}
where $I(z,t)$ and $U(z,t)$, respectively denote the intensity of
the current and the voltage at time $t$ and position $z$ of a finite
or infinite branch. Let us index these functions as
$\{L_j\}_{j=1}^{n+m}$, $\{C_j\}_{j=1}^{n+m}$, $\{I_j\}_{j=1}^{n+m}$
and $\{U_j\}_{j=1}^{n+m}$, defined on the branches
$\{e_j\}_{j=1}^{n+m}$.

At the central node, the Kirchhoff rules imply
\begin{align}\label{eq:kirchhoff}
&U_i(0)=U_j(0),\qquad i,j\in\{1,\cdots, m+n\}\notag\\
&\sum_{j=1}^{m+n} I_j(0)=0.
\end{align}
At the terminal nodes $l_{m+j}$ of the finite branch $e_{m+j}$, we
assume the boundary condition
\begin{equation}\label{eq:bc}
I_{m+j}(l_{m+j})=0 \qquad j\in\{1,\cdots, n\}.
\end{equation}

\subsection{Liouville transformation}
Here, we follow an old result by I. Kay~\cite{kay-72} and extended
by M. Jaulent to the dissipative case~\cite{Jaulent-82}.

For a wave of frequency $k$, i.e. for
\begin{align*}
I(z,t) &= I(k,z)e^{-\imath kt}\\
U(z,t) &= U(k,z)e ^{-\imath kt},
\end{align*}
the equation~\eqref{Jaulent1} becomes
\begin{equation}\label{quasiSchro}
\frac{d}{dz}\left(\frac{1}{L(z)}\frac{dU}{dz}\right)+k^{2}C(z)U=0
\end{equation}
Applying the Liouville transforation
$$
 x(z)=\int_{0}^{z}\left(L(u)C(u)\right)^{\frac{1}{2}} du
$$
and the conventions $I(k,z(x))=I(k,x)$, $U(k,z(x))=U(k,x)$, etc, and
setting
\begin{equation}\label{Jaulent3}
y(k,x)=\left[ \frac{C(x)}{L(x)}\right]^{\frac{1}{4}}U(k,x)
\end{equation}
the equation~\eqref{quasiSchro} writes
\begin{equation*}
\frac{d^{2}y}{dx^{2}}+(k^{2}-V(x))y=0.
\end{equation*}
Here, the potential $V(x)$ is defined as follows:
\begin{equation}\label{eq:potential}
V(x)=\left[
\frac{C(x)}{L(x)}\right]^{-\frac{1}{4}}\frac{d^{2}}{dx^{2}}\left[
\frac{C(x)}{L(x)}\right]^{\frac{1}{4}}.
\end{equation}
Note that, the new coordinate $x$ here has the dimension of time. It
actually parameterizes the travelling time of the wave through the
transmission line. In particular the length $l$ of a finite
transmission line is transformed into the correspoindg wave
travelling time:
$$
\tau=\int_0^{l} (L(z)C(z))^{1/2} dz.
$$
\subsection{Extension to the network}

The $LC$-transmission line equations defined on each edge for
$j\in\{ 1,\ldots,n+m\}$ are equivalent to the Schr\"odinger
equations
\begin{equation}\label{Schro}
\frac{d^{2}y_{j}}{dx^{2}}+(k^{2}-V_{j}(x))y_{j}=0.
\end{equation}
In this new formalism, the boundary condition~\eqref{eq:kirchhoff}
writes
\begin{align}\label{eq:kirchhoff2}
A_{j}^{-1}y_{j}\Big|_{x=0}&=A_{i}^{-1}y_{i}\Big|_{x=0}=\bar y\qquad
i,j\in\{1,\cdots,m+n\},\notag\\
\sum_{j=1}^{m+n}A_j y_j'\Big|_{x=0}&=\sum_{j=1}^{m+n}A_j'
y_j\Big|_{x=0}.
\end{align}
Here, applying the assumptions of the Section~\ref{sec:intro}, the
functions
$$
A_j(x_j)=\left[
\frac{C_j(x_j)}{L_j(x_j)}\right]^{\frac{1}{4}},\qquad
j\in\{1,\cdots, n+m\}
$$
are well defined, strictly positive and regular.

Furthermore, the boundary conditions~\eqref{eq:bc} write\small
\begin{equation}\label{eq:bc2}
y_{m+j}'(\tau_{m+j})=\frac{A_{m+j}'}{A_{m+j}}\Big|_{x=\tau_{m+j}}
y_{m+j}(\tau_{m+j})\qquad j\in\{1,\cdots, n\},
\end{equation}\normalsize
where $\{\tau_{m+j}\}_{j=1}^n$ denote the wave travelling time
corresponding the finite transmission lines.

In this paper, we denote
\begin{equation}\label{eq:defh}
h_{m+j}:=\frac{A_{m+j}'}{A_{m+j}}\Big|_{x=\tau_{m+j}}\qquad
j=1,\cdots,n.
\end{equation}

In conclusion, in order to study the $LC$-transmission line
equations on the graph $\Gamma$ with boundary
conditions~\eqref{eq:kirchhoff} and~\eqref{eq:bc}, we can study the
Schr\"odinger equation~\eqref{Schro} with boundary
conditions~\eqref{eq:kirchhoff2} and~\eqref{eq:bc2}.

\subsection{Self-adjointness}

We are dealing here with a symmetric operator $\LLL$ defined on
$L^2(\Gamma)$:\small
$$
\LLL=\oplus \LLL_j,\qquad \LLL_j=-\frac{d^2}{dx_j^2} + V(x_j) \text{
on } e_j, \quad j=1,\cdots,m+n,
$$\normalsize
with
$$
D(\LLL)= \text{closure of } C_{\text{b.c.}}^\infty \text{ in }
H^2(\Gamma).
$$
Here,  $C_{\text{b.c.}}^\infty$ denotes the space of infinitely
differentiable functions satisfying the boundary
conditions~\eqref{eq:kirchhoff2} and~\eqref{eq:bc2}.

In order to start the study of the scattering theory for this
operator, we need first to prove that it admits a self-adjoint
extension. In this aim, we assume an appropriate decay assumption on
$V(x_j)$ ensuring that the operator $\LLL$ is a compact perturbation
of the operator $\oplus_{j=1}^{m+n}
\left(-\frac{d^2}{dx_j^2}\right)$. Indeed, one also need to add a
new assumption \textbf{A4} to the three assumption \textbf{A1}
through \textbf{A3} of the introduction:
\begin{description}
  \item[A4] On each branch $e_j$, $j=1,\cdots,m+n$, we have
  $$
  \int_{0}^{\tau_j}(1+x_j)\Big|\left[\frac{C_j(x_j)}{L_j(x_j)}\right]^{\frac{1}{4}}\frac{d^{2}}{dx_j^{2}}\left[
\frac{C_j(x_j)}{L_j(x_j)}\right]^{-\frac{1}{4}} \Big|<\infty.
  $$
\end{description}
Now, we apply a general result by Carlson~\cite{carlson-98} on the
self-adjointness of differential operators on graphs.

Indeed, following the Theorem 3.4 of~\cite{carlson-98}, we only need
to show that at a node connecting $N$ edges, we have $N$ linearly
independent linear boundary conditions. At the terminal edges of
$\{e_j\}_{j=m+1}^{m+n}$ this is trivially the case as there is one
branch and one boundary condition. At the central edge it is not
hard to verify that~\eqref{eq:kirchhoff2} and~\eqref{eq:bc2} define
such boundary conditions as well. This implies that the operator
$\LLL$ is essentially self-adjoint and therefore that it admits a
unique self-adjoint extension. We are now ready to start the study
of the scattering solutions.

\section{Scattering Solutions}\label{sec:scat}

We are interested in the scattering solution where a signal of
frequency $k$ is applied at one end of one of the infinite branches.
In fact, in general, one might be interested in a experiment where
signals of different amplitudes $\{\alpha_j\}_{j=1}^m$ are applied
at all ends of the infinite branches. In such a case, we will be
seeking a scattering solution satisfying the asymptotic behavior at
$k\rightarrow\infty$
$$
y_j(x_j,k) \sim \alpha_j e^{-\imath k x_j}+R_j(k) e^{\imath k
x_j},\qquad j=1,\cdots,m.
$$
However, the experiment we consider here corresponds to the simple
case
$$
\alpha_1=1\quad \text{and} \quad \alpha_j=0,\qquad j=2,\cdots,m.
$$
The main result of this section, may be expressed as follows:
\begin{thm}\label{thm:scattering}
There exists a unique solution
$$
\Psi(x,k)=\left(y_j(x_j,k)\right)_{j=1}^{m+n},
$$
of the scattering problem, satisfying
\begin{itemize}
    \item $ \LLL_j y_j(x_j,k)=k^2 y_j(x_j,k)$;
    \item $\left(y_j(x_j,k)\right)_{j=1}^{m+n}$ satisfy the boundary
    conditions~\eqref{eq:kirchhoff2} and~\eqref{eq:bc2};
    \item For each $k\in\RR$, there exist $R_1(k)$ and
    $\{T_j(k)\}_{j=2}^k$ such that
    \begin{alignat}{2}
    y_1(x_1,k) \sim e^{-\imath k x_1}+R_1(k) e^{\imath k x_1},\quad
    x_1\rightarrow \infty,\label{eq:req1}\\
    y_j(x_j,k) \sim T_j(k) e^{\imath k x_j},\quad x_j\rightarrow
    \infty,\quad j=2,\cdots,m.\label{eq:req2}
    \end{alignat}
\end{itemize}
The coefficients, $R_1(k)$ and $\{T_j\}_{j=2}^m$, in~\eqref{eq:req1}
and~\eqref{eq:req2} appear to be unique. They are called
respectively the reflection and the transmission coefficients.
\end{thm}

We proceed the proof of the Theorem~\ref{thm:scattering} by
constructing such a solution and by showing that such a construction
is unique.

In this aim, we apply two different types of solutions: 1- the
so-called Jost solutions on
half-lines~\cite{Gerasimenko-87,Gerasimenko-88} and 2- the
fundamental solution of the one-end Sturm-Liouville
equation~\cite{marchenko-book}.

\subsection{Jost solutions}
The Jost solutions $f(x,k;V)$ and $\tilde f(x,k;V)$ on the half line
$[0,\infty)$, where $V$ is a potential defined on $[0,\infty)$, are
the solutions of the Schr\"odinger integral equations
\begin{align}\label{eq:jost}
f(x,k;V) &=e^{\imath k x} - \int_{x}^{\infty}\frac{\sin
k(x-y)}{k} V(y)f(y,k;V)dy,\notag\\
\tilde f(x,k;V) &= e^{-\imath k x} + \int_{0}^{x}\frac{\sin
k(x-y)}{k} V(y)\tilde{f}(y,k;V)dy.
\end{align}
Under the decay assumption A4 on $V$, these Volterra type integral
equations admit unique solutions. In fact, one only needs to apply
the standard iterative methods for the proof of existence and
uniqueness of solutions to Volterra type integral equations (see
e.g.~\cite{reed-simon3}, pages 138-139, for a proof).

The functions $f(x,k;V)$ and $\tilde f(x,k;V)$ satisfy the the
Schr\"odinger equation
$$
-\frac{d^2}{dx^2} f +V(x) f=k^2 f,\qquad -\frac{d^2}{dx^2} \tilde f
+V(x) \tilde f=k^2 \tilde f
$$
and admit the asymptotic behaviors (at $x\rightarrow \infty$)
\begin{align}\label{eq:jostinf}
f(x,k;V) & = e^{\imath k x}+o(1),\notag\\
\tilde f(x,k;V) &= a(k;V)e^{-\imath k x}+b(k;V) e^{\imath k x}+o(1),
\end{align}
where
\begin{align*}
a(k;V) &=  1-\frac{1}{2\imath k} \int_0^\infty e^{\imath k x} V(x) \tilde f(x,k;V)d x,\\
b(k;V) &=  \frac{1}{2\imath k} \int_0^\infty e^{-\imath k x} V(x)
\tilde f(x,k;V)d x.
\end{align*}
Moreover, one can show that~(\cite{reed-simon3}, pages 138-139)
\begin{equation}\label{eq:limder}
|\frac{d}{dx}f(x,k;V)-\imath k e^{\imath k x}|\leq
|k|\Big|\exp(\frac{1}{k}\int_x^\infty |V(y)|dy)-1\Big|.
\end{equation}
\subsection{Fundamental solution}

The fundamental solution, $\omega(x,k;h,V)$, is a solution of the
Schr\"odinger equation,
\begin{align}\label{eq:fund}
-\frac{d^2}{dx^2}\omega+V(x)\omega
&=k^2\omega,\qquad x\in(0,\infty),\notag\\
\omega(0,k;h,V)=1\quad &, \quad \omega'(0,k;h,V)=h,
\end{align}
where $\omega'$ denotes the derivative with respect to $x$.

This solution may be expressed through the following integral
representation~\cite{marchenko-book},
\begin{multline}\label{eq:fundint}
\omega(x,k;h,V)=\cos(kx)+h\frac{\sin(kx)}{k}\\+\int_{-x}^x
K(x,t;V)\{\cos(kt)+h\frac{\sin(k t)}{k}\}dt.
\end{multline}
Here the kernel $K(x,t;V)$ is a solution of the following integral
equation
\begin{multline}\label{eq:ker}
K(x,t;V)=\frac{1}{2}\int_{0}^{\frac{x+t}{2}}V(u)du\\
+\frac{1}{2}\int_{0}^{\frac{x+t}{2}}d\alpha\int_0^{\frac{x-t}{2}}
V(\alpha+\beta)K(\alpha+\beta,\alpha-\beta)d\beta,
\end{multline}
where we have extended the potential $V$ to be zero outside the
interval of its definition.

The integral equation~\eqref{eq:ker} admits a unique solution
(cf.~\cite{marchenko-book}, Theorem~1.2.2). Applying the same
result, for a potential $V$ in $L^1(\RR)$, this solution satisfies
\begin{equation}\label{eq:ineqker}
|K(x,t;V)|\leq \frac{1}{2}\|V\|_{L^1}\exp(x\|V\|_{L^1}),\quad
\text{for }x\geq 0 \text{ and } |t|\leq x.
\end{equation}
Furthermore, simple computations imply\small
\begin{multline*}
K_x(x,t;V)=\frac{1}{4}V(\frac{x+t}{2})\\+\frac{1}{4}\int_0^{\frac{x-t}{2}}d\beta~
V(\beta+\frac{x+t}{2})K(\frac{x+t}{2}+\beta,\frac{x+t}{2}-\beta)\\
+\frac{1}{4}\int_0^{\frac{x+t}{2}}d\alpha~
V(\alpha+\frac{x-t}{2})K(\alpha+\frac{x-t}{2},\alpha-\frac{x-t}{2}),
\end{multline*}\normalsize
and\small
\begin{multline*}
K_t(x,t;V)=\frac{1}{4}V(\frac{x+t}{2})\\+\frac{1}{4}\int_0^{\frac{x-t}{2}}d\beta~
V(\beta+\frac{x+t}{2})K(\frac{x+t}{2}+\beta,\frac{x+t}{2}-\beta)\\
-\frac{1}{4}\int_0^{\frac{x+t}{2}}d\alpha~
V(\alpha+\frac{x-t}{2})K(\alpha+\frac{x-t}{2},\alpha-\frac{x-t}{2}),
\end{multline*}\normalsize
where $K_x$ and $K_t$ denote respectively the derivatives with
respect to the first and the second coordinates $x$ and $t$. For a
potential $V\in L^1(\RR)$, this together with~\eqref{eq:ineqker}
implies
\begin{multline}\label{eq:ineqkerder}
|K_x(x,t;V)|,|K_t(x,t;V)|\leq
\frac{1}{4}|V(\frac{x+t}{2})|\\+\frac{1}{2}\|V\|^2_{L^1}\exp(x\|V\|_{L^1}),\quad
\text{for }x\geq 0 \text{ and } |t|\leq x.
\end{multline}

\subsection{Proof of Theorem~\ref{thm:scattering}: constructing the scattering solution}

For this construction, we distinguish between three types of edges.
\subsubsection{Edge $e_1$}

Define
$$
\zeta(x_1,k)=y_1(x_1,k)-\frac{1}{a(k;V_1)}\tilde f(x_1,k;V_1).
$$
Note that
\begin{equation}\label{eq:asya}
a(k;V)=1+\frac{1}{2\imath k}\int_{0}^\infty V(x)dx +o(1/|k|),
\end{equation}
and therefore, at least for $k$ large enough, $a(k,V_1)$ is non-zero
and $\zeta(x_1,k)$ is well-defined. Furthermore, note that
$\zeta(x_1,k)$ satisfies the Schr\"odinger equation
$$
-\frac{d^2}{dx_1^2}\zeta+V_1(x_1)\zeta=k^2 \zeta,
$$
and through the requirement~\eqref{eq:req1} and
by~\eqref{eq:jostinf}, at $x_1\rightarrow \infty$
\begin{equation}\label{eq:req3}
\zeta(x_1,k)\sim\left(R_1(k)t-\frac{b(k;V_1)}{a(k;V_1)}\right)e^{\imath
k x_1}.
\end{equation}
Consider now the Wronskian
\begin{multline*}
W(\zeta(.,k),\tilde f(.,k;V_1))(x_1)=\\
\zeta(x_1,k)\tilde f'(x_1,k;V_1)-\zeta'(x_1,k)\tilde f(x_1,k;V_1),
\end{multline*}
where the derivatives are taken with respect to the position $x_1$.
It is not a hard task to verify that the above Wronskian is a
constant function of $x_1$. In fact, as the both functions
$\zeta(.,k)$ and $\tilde f(.,k;V_1)$ verify the same Schr\"odinger
equation with the same potential $V_1$, one can easily check that
$$
\frac{d}{d x_1}W(\zeta(.,k),\tilde f(.,k;V_1))(x_1)= 0.
$$
Applying~\eqref{eq:req3} and~\eqref{eq:limder}, we have
$$
\lim_{x_1\rightarrow \infty} W(\zeta(.,k),\tilde f(.,k;V_1))(x_1) =0
$$
and therefore
$$
W(\zeta(.,k),\tilde f(.,k;V_1))(x_1)= 0\qquad \text{for
}x_1\in[0,\infty).
$$
This implies that $\zeta(.,k)$ and $f(.,k;V_1)$ are co-linear and
thus for some $R_1(k)$
\begin{multline}\label{eq:res1}
y_1(x_1,k)=\frac{1}{a(k;V_1)}\tilde
f(x_1,k;V_1)\\+(R_1(k)-\frac{b(k;V_1)}{a(k;V_1)})f(x_1,k;V_1).
\end{multline}

\subsubsection{Edges $e_j$, $j=2,\cdots,m$}

We consider the Wronskian $W(y_j(.,k),f(.,k;V_j))(x_j)$. Just as in
above, this Wronskian is a constant of $x_j$ and
applying~\eqref{eq:req2} together with~\eqref{eq:limder},
$$
\lim_{x_j\rightarrow \infty} W(y_j(.,k),\tilde f(.,k;V_j))(x_j) =0
$$
and thus for some $T_j(k)$
\begin{equation}\label{eq:res2}
y_j(x_j,k)=T_j(k)f(x_j,k;V_j),\qquad j=2,\cdots,m.
\end{equation}

\subsubsection{Edges $e_j$, $j=m+1,\cdots,m+n$}

We consider the Wronskian
$$
W(y_j(\tau_j-.,k),\omega(\tau_j-.,k;h_j,V_j(\tau_j-.)))(x_j),
$$
where $h_j$ is defined in~\eqref{eq:defh}. On the segment
$[0,\tau_j]$, the functions $y_j(\tau_j-.,k)$ and
$\omega(\tau_j-.,k;h_j,V_j(\tau_j-.))$ satisfy the same
Schr\"odinger equation with the same potential $V_j(\tau_j-.)$, and
therefore the Wronskian is a constant.

Applying~\eqref{eq:bc2} together with the definition of the
fundamental solution~\eqref{eq:fund}, we have
$$
W(y_j(.,k),\omega(\tau_j-.,k;h_j,V_j(\tau_j-.)))\Big|_{x_j=\tau_j}=0,
$$
and therefore, just as in above there exists a constant
$\alpha_j(k)$ such that
\begin{multline}\label{eq:res3}
y_j(x_j,k)=\alpha_j(k)\omega(\tau_j-x_j,k;h_j,V_j(\tau_j-.)),\\
j=m+1,\cdots,m+n.
\end{multline}
At this point, we can apply the boundary conditions at the centeral
node~\eqref{eq:kirchhoff2}, in order to find the $m+n$ constants
$R_1(k)$, $\{T_j(k)\}_{j=2}^m$ and $\{\alpha_j(k)\}_{j=m+1}^{m+n}$.
In fact, we have
\begin{align}\label{eq:bc3}
\bar y &= A_1^{-1}(0)
\left(\frac{1}{a(k;V_1)}+(R_1(k)-\frac{b(k;V_1)}{a(k;V_1)})f(0,k;V_1)\right)\notag\\
&=A_{j_1}^{-1}(0) T_{j_1}(k)f(0,k;V_{j_1})\notag\\
&=A_{j_2}^{-1}(0)
\alpha_{j_2}(k)\omega(\tau_j,k;h_j,V_{j_2}(\tau_j-.)),
\end{align}
for $j_1=2,\cdots,m$ and $j_2=m+1,\cdots,m+n$. Here, for the first
equality, we have applied $\tilde f(0,k;V_1)=1$. Moreover, we
have\small
\begin{multline}\label{eq:bc4}
A_1(0)(\frac{-\imath k}{a(k;V_1)}
+(R_1(k)-\frac{b(k;V_1)}{a(k;V_1)})f'(0,k;V_1))\\
+\sum_{j=2}^m A_j(0) T_j(k)f'(0,k;V_j)\\
+\sum_{j=m+1}^{m+n} A_j(0)
\alpha_j(k)\omega'(\tau_j,k;h_j,V_j(\tau_j-.)) =\bar
y\sum_{j=1}^{m+n} A_j(0)A_j'(0).
\end{multline}\normalsize
Here, for the first line we have applied $\tilde f'(0,k;V)=\imath
k$. One can easily check that,~\eqref{eq:bc3} and~\eqref{eq:bc4}
provide $m+n$ linearly independent linear equations for the $m+n$
unknown constants $R_1(k)$, $\{T_j(k)\}_{k=2}^m$ and
$\{\alpha_j\}_{k=m+1}^n$. One can therefore find these coefficients
uniquely and solve the scattering problem. $\square$

Here, we are only interested in the high frequency ($k\rightarrow
\infty$) behavior of the reflection coefficient $R_1(k)$ and we do
not need to compute all the coefficients explicitly.

\section{High-frequency asymptotic behavior of reflection coefficient}\label{sec:asy}

In this Section, we announce and prove the main result of this
paper:
\begin{thm}\label{thm:highfreq}
Consider the assumptions \textbf{A1} through \textbf{A4}, together
with the new one,
\begin{description}
  \item[A5] The function $L/C$ is continous at the central node od
  $\Gamma$; i.e.
  $$
  \frac{L_{j_1}(0)}{C_{j_1}(0)}=
  \frac{L_{j_2}(0)}{C_{j_2}(0)},\qquad j_1,j_2=1,\cdots,m+n.
  $$
\end{description}
Then, in the high-frequency regime ($k\rightarrow \infty$) the
reflection coefficient $R_1(k)$ satisfy:
\begin{equation}\label{eq:main}
R_1(k)\sim -\frac{(m-2)\imath -\sum_{j=m+1}^{m+n}\tan(k
\tau_j)}{m\imath-\sum_{j=m+1}^{m+n}\tan(k \tau_j)}+o(1).
\end{equation}
\end{thm}

Note that, applying this result one can easily identify $m$ the
number of the infinite branches, as well as
$\{\tau_j\}_{j=m+1}^{m+n}$ the wave travelling times of the finite
ones (when they are all different) through the reflection
coefficient $R_1(k)$. Indeed, through simple computations, we have
the main result of this paper:
\begin{cor}\label{cor:main}
Under the assumptions \textbf{A1} through \textbf{A5}, and in the
high frequency regime $k\rightarrow\infty$, we have
\begin{equation}\label{eq:m}
m=2\Re(\frac{1}{1+R_1(k)})+o(1),
\end{equation}
and
\begin{equation}\label{eq:tau}
\sum_{j=m+1}^{m+n}\tan(k\tau_j)=2\Im(\frac{1}{1+R_1(k)})+o(1),
\end{equation}
where $\Re$ and $\Im$ denote respectively the real and imaginary
parts of a complex number. In particular, scanning a certain
interval of frequencies $k$ for $k>k^*$ large enough, and through
the study of the poles of $\Im(\frac{1}{1+R(k)})$, we can identify
the wave travelling times $\tau_j$.
\end{cor}

Before starting the proof of the Theorem~\ref{thm:highfreq}, we need
to prove two lemmas.

\begin{lem}\label{lem:jost}
Under the assumption \textbf{A4} on $V(x)$, we have
\begin{equation}\label{eq:weyljost}
\frac{f'(0,k;V)}{f(0,k;V)}\sim \imath k+O(1)\qquad \text{as }
k\rightarrow \infty .
\end{equation}
\end{lem}
\vspace{.3cm}

\textit{Proof of Lemma~\ref{lem:jost}:} Here, we just need to apply
the the following approximations on $f(x,k;V)$ and $f'(x,k;V)$
(cf.~\cite{reed-simon3}, pages 138-139, for a proof):\small
\begin{align}\label{eq:approxjost}
|f(x,k;V)-e^{\imath k x}| &\leq \Big|\exp(\frac{1}{|k|}\int_x^\infty
|V(s)|ds)-1\Big|,\notag\\
|\frac{df(x,k;V)}{dx}-\imath k e^{\imath k x}| &\leq |k|
\Big|\exp(\frac{1}{|k|}\int_x^\infty |V(s)|ds)-1\Big|.
\end{align}\normalsize
Applying the integrability of $V(x)$, the
relation~\eqref{eq:weyljost} can be obtained very easily. $\square$
\begin{lem}\label{lem:fund}
Under the assumption \textbf{A4} on $V(x)$ and for a fixed position
$x=\tau$, we have
\begin{equation}\label{eq:weylfund}
\frac{\omega'(\tau,k;h,V)}{\omega(\tau,k;h,V)}\sim
-k\tan(k\tau)+o(1)\qquad \text{as } k\rightarrow \infty .
\end{equation}
\end{lem}

\textit{Proof of Lemma~\ref{lem:fund}:} We apply here the
expression~\eqref{eq:fundint} of $\omega(x,k;h,V)$. Through the
inequality~\eqref{eq:ineqker}, we now that for a fixed $\tau$, the
kernel $K(\tau,t;V)$ is bounded and therefore we easily have
$$
\omega(\tau,k;h,V)=\cos(k\tau)+\int_{-\tau}^\tau
K(\tau,t;V)\cos(kt)dt +O(1/k).
$$
developing the integral by parts, we have
\begin{multline*}
\int_{-\tau}^\tau
K(\tau,t;V)\cos(kt)dt=\frac{1}{k}\sin(kt)K(\tau,t;V)\Big|_{t=-\tau}^{t=\tau}\\
-\frac{1}{k}\int_{-\tau}^\tau \sin(kt)K_t(\tau,t;V) dt
\end{multline*}
Now, applying the inequality~\eqref{eq:ineqkerder}, we have
\begin{multline*}
\int_{-\tau}^\tau K(\tau,t;V)\cos(kt)dt \leq
\frac{1}{k}\sin(kt)K(\tau,t;V)\Big|_{t=-\tau}^{t=\tau} \\
-\frac{1}{4k}\int_{-\tau}^\tau |V(\frac{x+t}{2})|
dt+\frac{c}{k}\int_{-\tau}^\tau |\sin(kt)| dt\leq O(\frac{1}{k}),
\end{multline*}
where we have applied the boundedness of $K(\tau,t;V)$ for $|t|\leq
\tau$ and the integrability of $V$. This implies
\begin{equation}\label{eq:omegahigh}
\omega(\tau,k;h,V)=\cos(k\tau)+O(1/k), \quad \text{ as }
k\rightarrow\infty.
\end{equation}
Furthermore, we have\small
\begin{multline*}
\omega_x(\tau,k;h,V)=-k \sin(k\tau)+h\cos(k\tau)\\
+\left(K(\tau,\tau)+K(\tau,-\tau)\right)\cos(k
\tau)+h\left(K(\tau,\tau)-K(\tau,-\tau)\right)\frac{\sin(kt)}{k}\\
+\int_{-\tau}^\tau K_x(\tau,t;V)\{\cos (kt)+h\frac{\sin(kt)}{k}\}dt.
\end{multline*}\normalsize
Similar computations as in above, together with the
inequality~\eqref{eq:ineqkerder}, imply
\begin{equation}\label{eq:omegaderhigh}
\omega_x(\tau,k;h,V)=-k\sin(k\tau)+O(1), \quad \text{ as }
k\rightarrow\infty.
\end{equation}
The two relations~\eqref{eq:omegahigh} and~\eqref{eq:omegaderhigh}
finish the proof of Lemma~\ref{lem:fund} and we
have~\eqref{eq:weylfund}. $\square$

We are now ready to prove the Theorem~\ref{thm:highfreq}.

\textit{Proof of Theorem~\ref{thm:highfreq}:}

The assumption \textbf{A5} implies the existence of a constant $\bar
A$, such that
$$
A_{j}(0)=\bar A,\qquad \forall j=1,\cdots,m+n.
$$
Dividing~\eqref{eq:bc4} by $\bar y$ and applying~\eqref{eq:bc3}, we
have\small
\begin{align*}
\sum_{j=1}^{n+m}A_j'(0)/\bar A &= \frac{-\imath k
+(a(k;V_1)R_1(k)-b(k;V_1))f'(0,k;V_1)}{1+(a(k;V_1)R_1(k)-b(k;V_1))f(0,k;V_1)}
\\
&+\sum_{j=2}^{m}\frac{f'(0,k;V_j)}{f(0,k;V_j)}+ \sum_{j=m+1}^{m+n}
\frac{\omega'(\tau_j,k;h_j,V_j(\tau_j-.))}{\omega(\tau_j,k;h_j,V_j(\tau_j-.))}
.
\end{align*}\normalsize
Through this relation we can derive the value of the reflection
coefficient $R_1(k)$. The high-frequency behavior of this reflection
coefficient may be derived through the Lemmas~\ref{lem:jost}
and~\ref{lem:fund} and the relations~\eqref{eq:asya} and
\begin{equation}\label{eq:asyb}
b(k;V)\leq \|V\|_{L^1} O(1/k)\quad \text{ as } k\rightarrow \infty.
\end{equation}
Some simple computations, then imply the relation~\eqref{eq:main}
and finish the proof of the Theorem~\ref{thm:highfreq}. $\square$

\section{Conclusion}
As we have seen, the high-frequency asymptotic behavior of the
reflection coefficient, over a fixed semi-infinite edge of a
star-shape graph, must provide us very interesting and useful
information on the geometry and the topology of our network. Two
main directions may be considered for further extension of this
result. The first direction deals with the more general case of
trees instead of simply star-shape graphs. The second one
corresponds to the case of LCRG transmission lines where one needs
to extend all the utilities to the Zakharov-Shabat
operator~\cite{Jaulent-82}.

\bibliographystyle{plain}

\end{document}